\documentclass[12pt]{article}
\usepackage{graphicx}

\def\kek{High Energy Accelerator Research Organization\\
Tsukuba, JAPAN}

\def\Title#1{\begin{center} {\Large #1 } \end{center}}
\def\Author#1{\begin{center}{ \sc #1} \end{center}}
\def\Address#1{\begin{center}{ \it #1} \end{center}}

\newenvironment{Abstract}{\begin{quotation}  }{\end{quotation}}
\newenvironment{Presented}{\begin{quotation} \begin{center}
             PRESENTED AT\end{center}\bigskip
      \begin{center}\begin{large}}{\end{large}\end{center} \end{quotation}}

\def\beq{\begin{equation}}
\def\eeq#1{\label{#1}\end{equation}}
\def\eeqn{\end{equation}}

\def\beqa{\begin{eqnarray}}
\def\eeqa#1{\label{#1}\end{eqnarray}}
\def\eeqan{\end{eqnarray}}

\let\bar=\overbar

\def\Dslash{\not{\hbox{\kern-4pt $D$}}}
\def\dslash{\not{\hbox{\kern-2pt $\del$}}}

\def\msb{{\bar{\ssstyle M \kern -1pt S}}}

\begin{document}
\begin{titlepage}

\vfill
\Title{Experimental Prospects for $B \to X_{s/d}\gamma$
and $B \to X_s\ell^+\ell^-$}
\vfill
\Author{Shohei Nishida}
\Address{\kek}
\vfill
\begin{Abstract}
 In this report, experimental prospects for the inclusive analysis
 of the radiative $B$ decays and electroweak penguin decays
 at the super $B$ factories are presented.
\end{Abstract}
\vfill
\begin{Presented}
Proceedings of CKM2010,
the 6th International Workshop on the CKM Unitarity Triangle, \\
University of Warwick, UK, 6-10 September 2010
\end{Presented}
\vfill
\end{titlepage}
\def\thefootnote{\fnsymbol{footnote}}
\setcounter{footnote}{0}
%

\newcommand{\abi}{\mathrm{ab}^{-1}}
\newcommand{\ACP}{A_{CP}}
\newcommand{\AFB}{A_{\mathrm{FB}}}
\newcommand{\BB}{B\bar{B}}
\newcommand{\BF}{\mathcal{B}}
\newcommand{\bsgamma}{b \to s\gamma}
\newcommand{\bsll}{b \to s\ell^+\ell^-}
\newcommand{\CP}{\mathit{CP}}
\newcommand{\DE}{\Delta E}
\newcommand{\Egamma}{E_\gamma}
\newcommand{\ellell}{\ell^+\ell^-}
\newcommand{\fbi}{\mathrm{fb}^{-1}}
\newcommand{\KS}{K_S^0}
\newcommand{\GeV}{\mathrm{GeV}}
\newcommand{\Mbc}{M_{\mathrm{bc}}}
\newcommand{\MeV}{\mathrm{MeV}}
\newcommand{\VtdVts}{V_{td}/V_{ts}}
\newcommand{\Xd}{X_d}
\newcommand{\Xsd}{X_{s/d}}

\newcommand{\PM}[2]{{\,}^{+#1}_{-#2}}

\section{Introduction}

Radiative $B$ decay $B \to X_s\gamma$, $X_d\gamma$ and
electroweak penguin decay $B \to X_s\ellell$,
where $X_s$ ($X_d$) is the hadronic recoil system including
an $s$ ($d$) quarks, are the flavor changing neutral processes,
sensitive to New Physics (NP).
After the first measurement of the inclusive branching fraction
of $B \to X_s\gamma$ by CLEO~\cite{Alam:1994aw},
several measurements of inclusive
$B \to \Xsd\gamma$ and $B \to X_s\ellell$ have been performed
by BaBar and Belle.

In general, the inclusive measurements are experimentally challenging,
but theoretically clean.
Therefore, improving existent measurements and
exploring new measurements
of the inclusive $B \to \Xsd\gamma$, $X_s\ellell$ processes
with much larger luminosity at the super $B$ factories
will provide important test of NP.
In this proceedings, experimental prospects of the measurements
of branching fractions and other observables, such as $\CP$
asymmetry, of the inclusive $B \to \Xsd\gamma$, $X_s\ellell$
are reported.


\section{Branching fraction of $B \to X_s\gamma$}

The experimental analysis of the inclusive $B \to X_s\gamma$ decay
has been performed with three methods: (1) fully inclusive,
(2) sum of exclusive modes and (3) recoil tag.

In the fully inclusive method, we substract the on-resonance
photon energy spectrum by the continuum spectrum.
This method is free from the model uncertainty of the hadronic
recoil system $X_s$. However, it sufferes large backgrouds
from the continuum process and $B$ decays.
Therefore, a lepton from the other side $B$
is sometimes tagged for the continuum background suppression.
This lepton tag is also useful for flavor tagging.

In the method using sum of exclusive modes,
which are often refered to as semi-inclusive or pseudo-reconstruction
method, hadronic
system $X_s$ is reconstructed as a sum of exclusive final states.
This method provides higher purity than
the fully inclusive method, and clear separation between
$X_s$ and $X_d$. On the other hand, it suffers from large
model uncertainty of the hadronic system partially because of the
modes that are not reconstructed.

In the recoil tag method,
one $B$ meson is either fully reconstructed
with hadronic final states or
is tagged with semi-leptonic $B$ decay,
and a high energy photon coming from $B \to X_s\gamma$
is looked for among remaining particles in the event.
In this method, signal is very clean and continuum background is negligible.
The drawback is very low efficiency ($\sim O(0.1\%)$)
that requires huge amount of statistics.
When the tag side $B$ meson is fully reconstructed,
the photon energy in the $B$ rest frame and flavor information
can be obtained.

At present, the most precise measurement of $\BF(B \to X_s\gamma)$
is obtained with fully inclusive method.
Belle obtained
$\BF(B \to X_s\gamma) =
(3.45 \pm 0.15 \pm 0.40) \times 10^{-4}$
with $\Egamma > 1.7~\GeV$~\cite{:2009qg}.
The error is already systematic dominated.

Therefore, in the super $B$ factories with more than $10~\abi$,
the recoil tag method is most promising.
If we scale BaBar's result
$\BF(B \to X_s\gamma; \Egamma > 1.9~\GeV) =
(3.66 \pm 0.85 \pm 0.60) \times 10^{-4}$
with $210~\fbi$~\cite{Aubert:2007my},
the statistical error at $10~\abi$ will be 3\%.

\begin{figure}[htb]
 \centering
 \includegraphics[scale=0.7]{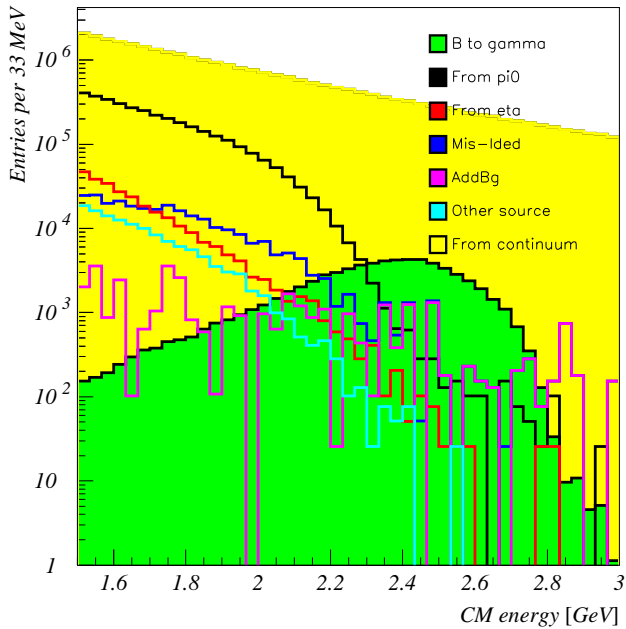}
 \caption{Breakdown of the background component of
 $B \to X_s\gamma$ analysis with the fully inclusive method.
 $B \to \gamma$ component is the $B \to X_s\gamma$ signal.
 The background from continuum will be less significant
 in the recoil tag method.}
 \label{fig:bsgamma-egam-b}
\end{figure}

The challenge will be to reduce the systematic errors.
Fig.~\ref{fig:bsgamma-egam-b} shows the breakdown of the background
in the fully inclusive decay.
In the recoil tag method, background from the continuum process
will be negligible, but other contribution is not expected to be
significantly smaller.
Among the background from $B$ decays,
decays of $\pi^0$ and $\eta$ are the major contrinbutions,
but they can be calibrated from control samples.
However, there are some components difficult to calibrate
such as decay of $\omega$, $\eta'$, $J/\psi$ or
hadronic interaction of neutral particle in the calorimeter.
According to the inclusive analysis by Belle, around $7\%$
systematic error is assigned to $B$ background except
$\pi^0$, $\eta$ decay.
These number can be reduced in future, but
the systematic error of $3$-$5\%$ is expected.
Nevertheless, this might be adequate
given the current theoretical prediction
$\BF(B \to X_s\gamma) = ( 3.15 \pm 0.23 ) \times 10^{-4}$~\cite{Misiak:2006zs}.

\section{$\CP$ Asymmetry of $B \to X_{s/d}\gamma$}

The $\CP$ asymmetry of $B \to X_{s/d}\gamma$ is predicted
very accurately by theory, and hence will be
a sensitive probe to NP in the super $B$ factories.

A prediction~\cite{Hurth:2003dk} shows
$\ACP(B \to X_s\gamma) = 0.0044 \PM{0.0024}{0.0014}$
and
$\ACP(B \to X_d\gamma) = - 0.102 \PM{0.033}{0.058}$,
while the $\CP$ asymmetry for the sum of $B \to X_s\gamma$
and $B \to X_d\gamma$ ($B \to X_{s+d}\gamma$) is essentially zero.
On the other hand, some NP models predict larger $\CP$
asymmetry for $B \to X_s\gamma$ and $B \to X_{s+d}\gamma$.
Therefore, measurement of these asymmetries provide useful information
to identify NP models.

The $\CP$ asymmetry in $B \to X_s\gamma$ has been performed
with analysis using sum of exclusive modes.
The advantage of this method is that the flavor information
can be obtained from the reconstruction information.
Also, the contribution of $B \to X_d\gamma$ is negligible.
Belle obtained $\ACP(B\to X_s\gamma) = 0.002 \pm 0.050 \pm 0.030$
with $140~\fbi$~\cite{Nishida:2003yw}
and BaBar obtained 
$\ACP(B\to X_s\gamma) = -0.011 \pm 0.030 \pm 0.014$ with a dataset
with $383$ M $\BB$~\cite{:2008gvb}.
Here, most of the systematic errors are limited by the control
sample statistics, and can be reduced in future.
If we scale the Belle result,
the expected statistical and systematic errors are
$\pm 0.009 \mathrm{(stat)} \pm 0.006 \mathrm{(syst)}$
at $5~\abi$ and
$\pm 0.003~\mathrm{(stat)} \pm 0.002~\mathrm{(syst)}
\pm 0.003~\mathrm{(model)}$~\cite{Aushev:2010bq}.

In the fully inclusive method, lepton tag is useful to
tag the flavor of $B$ meson, though wrong tag happens due
to $B$ mixing ($\sim 9\%$) and leptons from non-$B$ ($\sim 3\%$).
Since $B \to X_d\gamma$ is inevitably included in the signal,
this method is useful to measure the asymmetry for $B \to X_{s+d}\gamma$.
By simple extrapolation, $1\%$ statistical error is possible
at $10~\abi$.

\section{Branching fraction of $B \to X_d\gamma$}

Branching fraction of $B \to X_d\gamma$ is useful
to constrain $|\VtdVts|$.
However, the measurement suffers huge background from $B \to X_s\gamma$ decays.
In order to suppress contamination from $B \to X_s\gamma$,
analysis with sum of exclusive modes is the most promising.

According to the MC study by Belle~\cite{Aushev:2010bq},
one expects $5\%$ statistical errors at $5~\abi$
if we sum up 2 to 4 pions including up to 1 $\pi^0$ to
reconstruct $X_d$ up to $2.0~\GeV$.
However, the systematic error is around $20\%$, which mainly
comes from the normalization of $\bsgamma$ components.

BaBar performed the study of $B \to X_d\gamma$ with 
a dataset with $471$~M $\BB$~\cite{:2010ps}.
They reconstructed 7 exclusive modes at $0.5 < M_{X_{d(s)}} < 2.0~\GeV$
for $B \to X_d\gamma$ and $B \to X_s\gamma$,
and measured $\BF(B \to X_d\gamma) = (9.2 \pm 2.0 \pm 2.3) \times 10^{-6}$
and $\BF(B \to X_s\gamma) = (23.0 \pm 0.8 \pm 3.0)$ respectively,
obtaining $|\VtdVts| = 0.199 \pm 0.022 \pm 0.024 \pm 0.002 \mathrm{(th.)}$.
Again, the key issue in future is the reduction of the systematic error.
A large part of it comes from unreconstructed modes
and poor knowledge about the final states.
Systematic error can be reduced
 by adding more reconstruction modes and more statistics,
but will remain the dominant source of the error at the super $B$
factories.

Another possibility for the study of $B\to X_d\gamma$
is to use the recoil tag method and
apply strangeness tag to remove $B \to X_s\gamma$ component.
However, the strangeness tag is not straightforward because of
the neutral kaons, baryons and possible $s\bar{s}$ popping.
However the method can be only used with $50~\abi$
or more~\cite{Hewett:2004tv}.

\section{$B \to X_s\ellell$}

The inclusive analysis of $B \to X_s\ellell$ is more challenging
than $B \to \Xsd\gamma$ because of  two orders of magnitudes
lower branching fraction.
Similarly to exclusive analyses like $B \to K^*\ellell$,
there exist many observables to be measured 
in addition to the branching fraction,
such as $\CP$ asymmetry, forward-backward asymmetry.
The inclusive modes are theoretically clean compared to exclusive
modes.
For example, the zero-crossing point of the $q^2$ distribution
of the forward backward aymmetry $q_0^2$ is predicted to
$3.50 \pm 0.12~\GeV^2$ ($3.38 \pm 0.11~\GeV^2$)
for $B \to X_s\mu^+\mu^-$ ($X_se^+e^-$)~\cite{Huber:2007vv}
and $4.2 \pm 0.6~\GeV^2$ for $B \to K^*\ellell$~\cite{Feldmann:2002iw}.
Experimentally, the analysis is very challenging, but is possible
only in $e^+e^-$ $B$ factories.

\begin{figure}[htb]
 \centering
 \begin{tabular}{ll}
  (a) & (b) \hspace{38mm}(c) \\
  \includegraphics[scale=0.46]{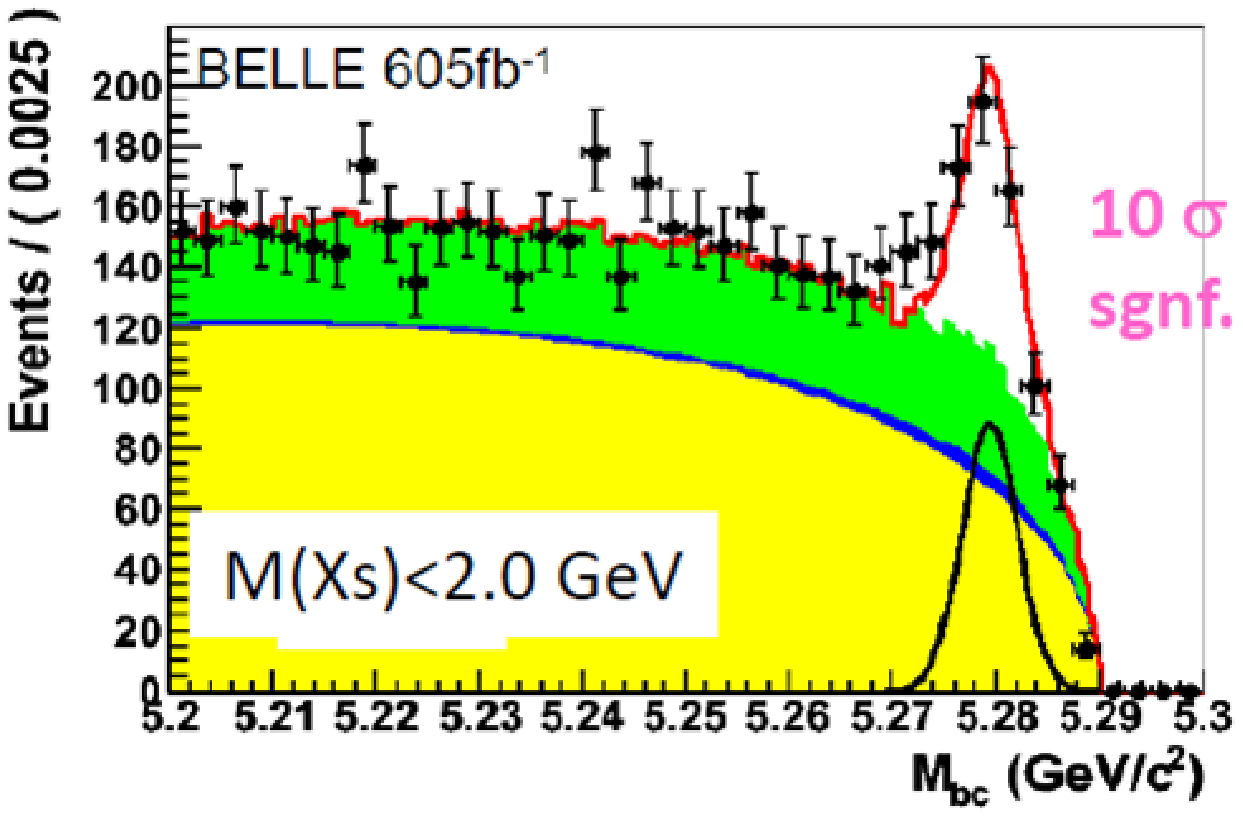}
  & \includegraphics[scale=0.46]{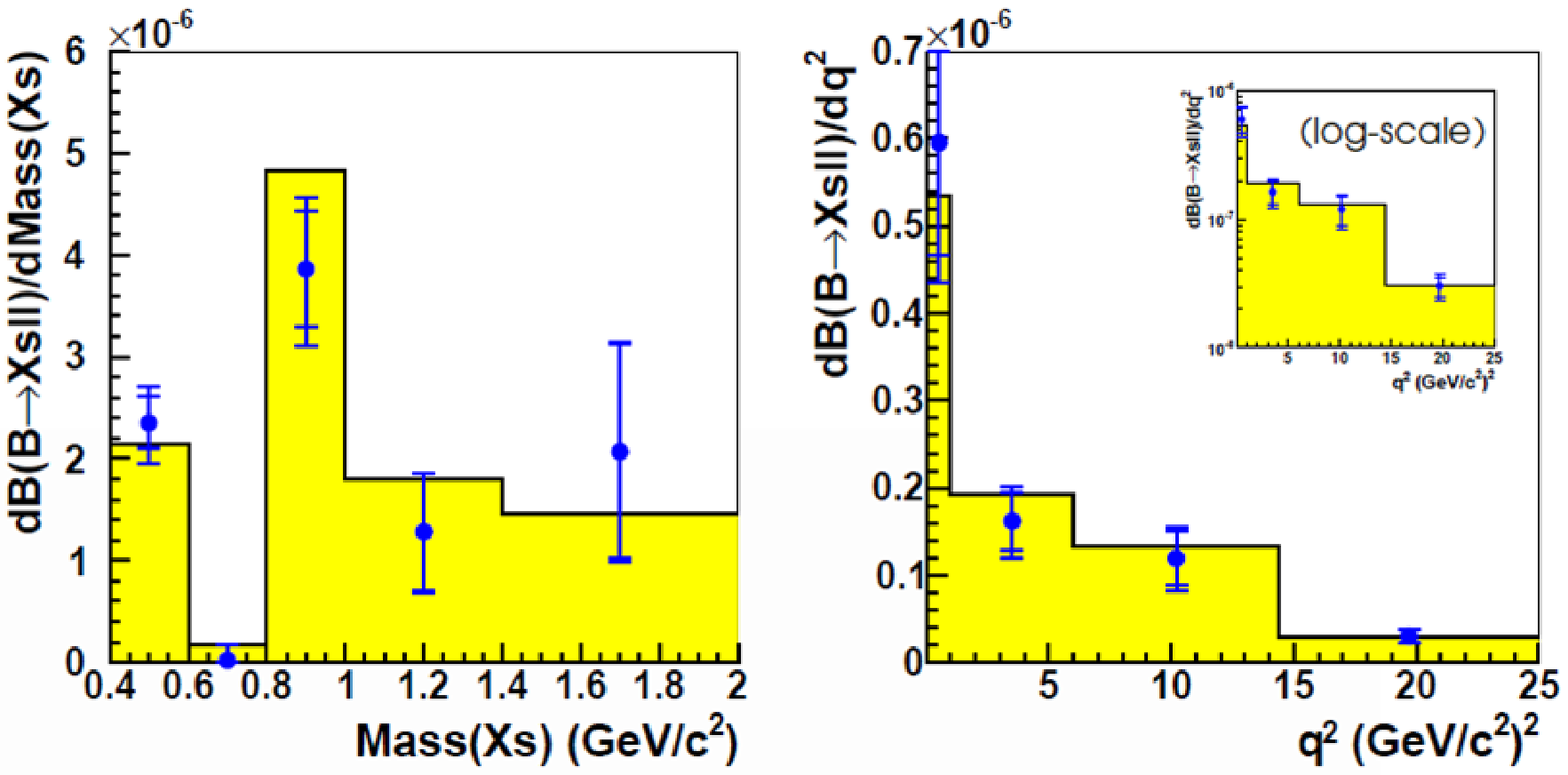}
 \end{tabular}
 \caption{Inclusive $B \to X_s\ellell$ analysis at Belle with
 $605~\fbi$. (a) $\Mbc$ distribution for $M_{X_s} < 2.0~\GeV$
 (b) $M_{X_s}$ dependence of the branching fraction
 (c) $q^2$ dependence of the branching fraction.}
 \label{fig:xsll}
\end{figure}

So far, all the analyses at BaBar and Belle are with
the sum of exclusive modes.
Figure~\ref{fig:xsll} shows the result of $B \to X_s\ellell$
by Belle with $605~\fbi$.
Belle observed $238.3 \pm 26.4 \pm 2.3$ events, resulting
the branching fraction of
$(3.33 \pm 0.80 \PM{0.19}{0.24}) \times 10^{-6}$.
The result is consistent with the SM prediction of
$(4.2 \pm 0.7) \times 10^{-6}$~\cite{Ali:2002jg}.
The $M_{X_s}$ and $q^2$ dependence for a few bins are obtained.
Although the error is still dominated by the statistical error
at present, it will be important to reduce the systematic error
at the super B factories.
According to the study for SuperB~\cite{O'Leary:2010af},
the statistical error can be reduced to a few percent with $75~\abi$.
One of the main source of the systematic
error is the uncertainty of the hadronic system 
and the unreconstructed modes, which may be 
reduced by comparing with $B \to X_s\gamma$ decays.

The next step is to measure the forward backward asymmetry
for $B \to X_s\ellell$. Unfortunately,
the sensitivity of the forward backward asymmetry is not estimated
yet for inclusive $B \to X_s\ellell$.
In the exclusive $B \to K^*\ellell$ mode, $q_0^2$
is expepected to be measurable with the precision of $5\%$ at $50~\abi$.
However, the precision for the inclusive modes is expected to be
significantly worse than for the exclusive modes.
Considering the theoretical precision,
forward backward asymmetry of the inclusive $B \to X_s\ellell$
is not competitive to exclusive modes at early stage of the super
$B$ factories.




\end{document}